\documentclass[aps,twocolumn,pre,superscriptaddress,showpacs,reprint]{revtex4-1}
\usepackage{times}
\usepackage{xspace}
\usepackage{graphicx}
\usepackage{dcolumn}
\usepackage{upgreek}
\usepackage{color}
\usepackage{amsmath, amsthm, amssymb}
\usepackage{bm}
\usepackage{wrapfig}

\newcommand{\ucsdphysics}{Department of Physics, University of California,
                          San Diego, La Jolla CA 92093}
\newcommand{\ucsdmath}{Department of Mathematics, University of California,
                          San Diego, La Jolla CA 92093}
\newcommand{\ctbp}{Center for Theoretical Biological Physics, University of California, San Diego}
\newcommand{\herbierice}{Department of Bioengineering, Center for Theoretical Biological Physics, Rice University, Houston, TX 77005}

\newcommand{\eq}{Eq.~}
\newcommand{\eqs}{Eqs.~}
\newcommand{\fig}{Fig.~}

\newcommand{\rhoac}{\ensuremath{m_0\xspace}}
\newcommand{\lh}{\ensuremath{\ell_h}\xspace}
\newcommand{\vb}[1]{{\bf #1}}
\newcommand{\de}{\ensuremath{\delta_{\epsilon}}\xspace}
\newcommand{\um}{\ensuremath{\upmu\textrm{m}}\xspace}

\newcommand{\rb}{\ensuremath{\vb{r}}\xspace}

\newcommand{\uv}{\vb{u}\xspace}

\newcommand{\spt}{\ensuremath{\sigma_{\textrm{poly}}}}
\newcommand{\lc}{\ensuremath{L_{\textrm{cell}}}\xspace}
\newcommand{\sm}{\ensuremath{\sigma_{\textrm{myo}}}}

\newcommand{\sech}{\ensuremath{\textrm{sech}}\xspace}

\begin{document}

\title{Periodic migration in a physical model of cells on micropatterns}
\author{Brian~A.~Camley}
\affiliation{\ucsdphysics}
\affiliation{\ctbp}
\author{Yanxiang Zhao}
\affiliation{\ucsdmath}
\affiliation{\ctbp}
\author{Bo Li}
\affiliation{\ucsdmath}
\affiliation{\ctbp}
\author{Herbert Levine}
\affiliation{\herbierice}
\author{Wouter-Jan Rappel}
\affiliation{\ucsdphysics}
\affiliation{\ctbp}

\begin{abstract}
We extend a model for the morphology and dynamics of a crawling
eukaryotic cell to describe cells on micropatterned substrates.  This
model couples cell morphology, adhesion, and cytoskeletal flow in
response to active stresses induced by actin and
myosin.  We propose that protrusive stresses are only generated where
the cell adheres, leading to the cell's effective
confinement to the pattern.  Consistent with experimental results, 
simulated cells exhibit a broad range
of behaviors, including steady motion, turning, bipedal motion, and
periodic migration, in which the cell crawls persistently in one
direction before reversing periodically.  We show that periodic
motion emerges naturally from the coupling of cell polarization to
cell shape by reducing the model to a simplified one-dimensional form
that can be understood analytically.  \end{abstract}

\maketitle

Cultured cells on two-dimensional substrates are often
used as a convenient proxy for more biologically relevant situations,
such as cells within three-dimensional extracellular matrix (ECM).
However, cells in ECM often exhibit qualitatively 
different modes of migration than those on substrates
\cite{doyleyamada,poincloux2011contractility,zaman2006migration,yamazaki2009involvement}.
A remarkable example of this is the discovery of periodic migration
in zyxin-depleted cells in collagen matrix
\cite{fraley2012dimensional}.  
Understanding cell motility in ECM may
be profoundly important for the study of cancer invasion
\cite{wirtz2011physics}.
Interestingly, features of cell morphology and dynamics in matrix are recapitulated in cells on micropatterned adhesive substrates, including cell speed, shape, dependence on myosin
\cite{doyleyamada} and periodic migration
\cite{fraley2012dimensional}.  Other micropatterns induce cell polarization and directed cell motion
\cite{thery_review,thery2006anisotropy,mahmud2009directing} and
sorting of cells from one- to two-dimensional regions of
micropatterns \cite{chang2013guidance}.  
In this 
Letter, we study the influence of
micropatterns on cell motility using an extension 
of a computational model of eukaryotic cell crawling \cite{shao2010computational,shao2012coupling} and 
observe 
a wide range of dynamic behaviors
including periodic migration.  To our knowledge, ours is the first cell crawling simulation to display periodic migration.

It would be natural to expect that periodic migration \cite{fraley2012dimensional} requires underlying
oscillatory protein dynamics, as in {Min oscillations in E. coli
\cite{lutkenhaus2007assembly}. }
Surprisingly, this
is not the case; periodic migration and other complex behaviors appear with only minimal alteration
to the model for freely crawling cells.  We study periodic migration in 
detail, and show that it is a consequence of feedback between the cell's shape and its biochemical polarization, {i.e. how proteins are segregated to one side of the cell}.  We use
sharp interface theory to reduce our model to a 
simplified one-dimensional (1D) model that is analytically tractable.  
Periodic migration exemplifies how
coupling between cell shape and chemical polarity can lead to unexpected cell behavior.

{\it Model summary.}  We describe the cell's cytoskeleton as a viscous, compressible fluid 
driven by active stresses from actin
polymerization and myosin contraction.   This is appropriate for the long time scales of keratocyte and fibroblast migration on which the cytoskeleton can rearrange, see e.g. \cite{rubinstein2009actin}.  {Our model is one of a broad spectrum of active matter \cite{juelicher2007active,marchetti2013hydrodynamics} models of motility in which active stresses drive deformation \cite{recho2013contraction,hawkins2009pushing,tjhung2012spontaneous,whitfield2013active,callan2013active,herant2010form,joanny2012drop,kruse2006contractility,voituriez2006generic,rubinstein2009actin,wolgemuth2011redundant}.}  Details of the model are available in Ref.  
\cite{shao2012coupling}; we review it briefly to
highlight changes made to study cells on micropatterns.  It has
four modules: 1) cell shape, tracked by a phase field
$\phi(\rb,t)$, 2) the cytoskeleton as an active viscous
compressible fluid \cite{rubinstein2009actin,bois2011pattern}, 3) actin promoter (e.g. Rac or Cdc42) and myosin concentrations obeying
reaction-diffusion-advection equations, and 4) adhesions
between cell and substrate, tracked individually.  

Cell shape is tracked by a ``phase field'' $\phi(\rb,t)$ that is
zero outside and unity inside the cell \cite{kockelkoren2003computational,boettinger2002phase,collins1985diffuse,shao2010computational,biben2005phase,ziebert2012model,li2009solving}. 
$\phi$ varies smoothly across the cell boundary, which is
implicitly set by $\phi = 1/2$.  $\phi(\rb,t)$ obeys \begin{equation}
\partial_t \phi + \uv \cdot \nabla \phi = \Gamma \left( \epsilon
\nabla^2 \phi -G'(\phi)/\epsilon + c \epsilon |\nabla \phi| \right)
\label{eq:phasefield}
\end{equation} where $\uv$ is the cytoskeletal velocity, $\Gamma$ a
relaxation coefficient, $c = \nabla\cdot\left(\frac{\nabla\phi}{|\nabla\phi|}\right)$ is the local interface curvature, $\epsilon$ the interface width, and $G(\phi) = 18\phi^2(1-\phi)^2$.

We describe cytoskeletal flow with a Stokes equation including active
forces from actin and myosin and forces induced by membrane
curvature and cell-substrate adhesion: 
\begin{align}
\nabla \cdot\left[\nu\left(\nabla \uv + \nabla \uv^T\right) \right] &+
\nabla \cdot (\spt + \sm) \label{eq:stokes}\\ \nonumber &+
\vb{F}_{\textrm{mem}} + \vb{F}_{\textrm{adh}} -\xi \uv= 0 
\end{align}
where $\nu(\phi) = \nu_0 \phi$ is the viscosity.  
$\xi$ does not vary over the substrate i.e. $\xi \uv$ is a hydrodynamic drag \cite{evans1988translational}, not friction from
adhesive binding \cite{walcott2010mechanical}. Individual adhesions lead to $\vb{F}_{\textrm{adh}}$; $\vb{F}_{\textrm{mem}}$ comes from membrane deformations (see Appendix).  
We neglect the pressure term arising from 
coupling between cytoskeletal mesh and cytoplasm \cite{rubinstein2009actin}.  \eq \ref{eq:stokes} is
solved numerically with a semi-implicit finite difference spectral method; other equations are stepped explicitly (see Appendix).

Our central hypothesis for the effect of the adhesive micropattern is
that protrusive stress from actin polymerization, \spt, is only generated where the
cell contacts the micropattern, 
\begin{equation} \spt = -\eta_a^0 \chi(\rb)
\phi \rho_a \de \hat{\bf{n}}{\hat{\bf{n}}} 
\end{equation}
where $\chi(\rb)$ is one inside the pattern and zero outside,
$\de(\phi) = \epsilon |\nabla \phi|^2$,
$\hat{\bf{n}}$ is the normal to the cell surface, $\rho_a$ the actin promoter density on the membrane, and $\eta_a^0$ a protrusion coefficient. Our assumption is 
supported by experimental work showing that 
fibroblasts preferentially protrude processes from points near newly
formed adhesions, which only form on the pattern \cite{xia2008directional}.  Others have proposed active stresses proportional to cell-substrate adhesion
\cite{carlsson2011mechanisms}.   Our pattern is a stripe, $\chi(\rb)
= \frac{1}{2}\left[1 + \tanh(3\left\{ \frac{w}{2}-|x|
\right\}/\epsilon )\right]$, with $w$ the stripe width.  The contractile stress is $\sm = \eta_m^0 \phi \rho_m \vb{I}$ with
$\rho_m$ the myosin density, $\eta_m^0$ the myosin contractility coefficient, and $\vb{I}$ the identity tensor.  

Cell polarization arises from $\rho_a$, which follows a
wave-pinning model \cite{mori2008wave}.  Actin promoter exchanges between
active membrane-bound ($\rho_a$) and inactive cytosolic ($\rho_a^{\textrm{cyt}}$) states; membrane-bound promoter catalyzes binding to the membrane.  Fronts between high $\rho_a$ and low $\rho_a$ can stall (``pin''), leading to a steady polarization \cite{mori2008wave}.

Actin promoter and myosin processes only occur
inside the cell; the phase field method is ideally suited 
to handle 
reaction-diffusion-advection equations within moving cells
\cite{shao2010computational,biben2005phase,ziebert2012model,li2009solving,tjhung2012spontaneous}. 
The reaction-diffusion-advection equations for {actin promoter} and myosin are
\begin{align}
\partial_t\left(\phi \rho_a\right) + \nabla \cdot \left(\phi \rho_a \uv \right) &= \nabla \cdot \left[\phi D_a \nabla \rho_a\right] + \phi f \\
\partial_t\left(\phi \rho_m\right) + \nabla \cdot \left(\phi \rho_m \uv \right) &= \nabla \cdot \left[\phi D_m(\rho_a) \nabla \rho_m\right]. \label{eq:rd_myo}
\end{align}
{Actin promoter diffuses with coefficient $D_a$ on the membrane; {at this level of modeling, we do not distinguish between membrane and cytoskeleton velocity, and so $\rho_a$ is advected with the cytoskeletal velocity \uv}.  Myosin binds and unbinds from the cytoskeleton, which we model as a $\rho_a$-dependent diffusion coefficient $D_m(\rho_a) = D_m^0/(1+\rho_a/K_D)$ \cite{shao2012coupling}.}
The nonlinear reaction term $f(\rho_a,\rho_a^{\textrm{cyt}})$ for promoter membrane-cytosol exchange is
in the Appendix.  $\rho_a^{\textrm{cyt}}$ is
well-mixed (constant) and set by the conservation of $\rho_a$, i.e.
$\int d^2 r \phi(\rb) \left[\rho_a(\rb) + \rho_a^{\textrm{cyt}}
\right] = N_a^{\textrm{tot}}$ is constant. 

Adhesions between cell and substrate are
formed, age, and transition between modes as in \cite{shao2012coupling}.  However, adhesions may
only form on the micropattern \cite{xia2008directional}; adhesions that leave the micropattern are destroyed (see Appendix).  The number of adhesions is fixed.  We do not enforce symmetry, unlike \cite{shao2012coupling}.  

\begin{figure}[ht!]
\includegraphics[width=85mm]{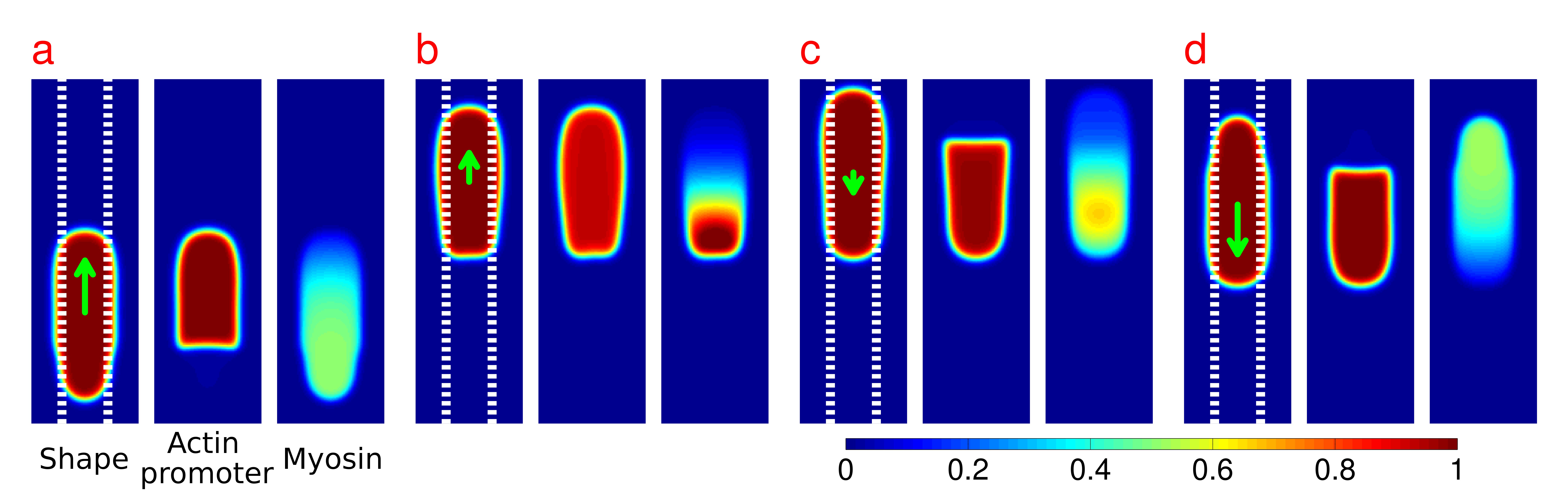}
\includegraphics[width=85mm]{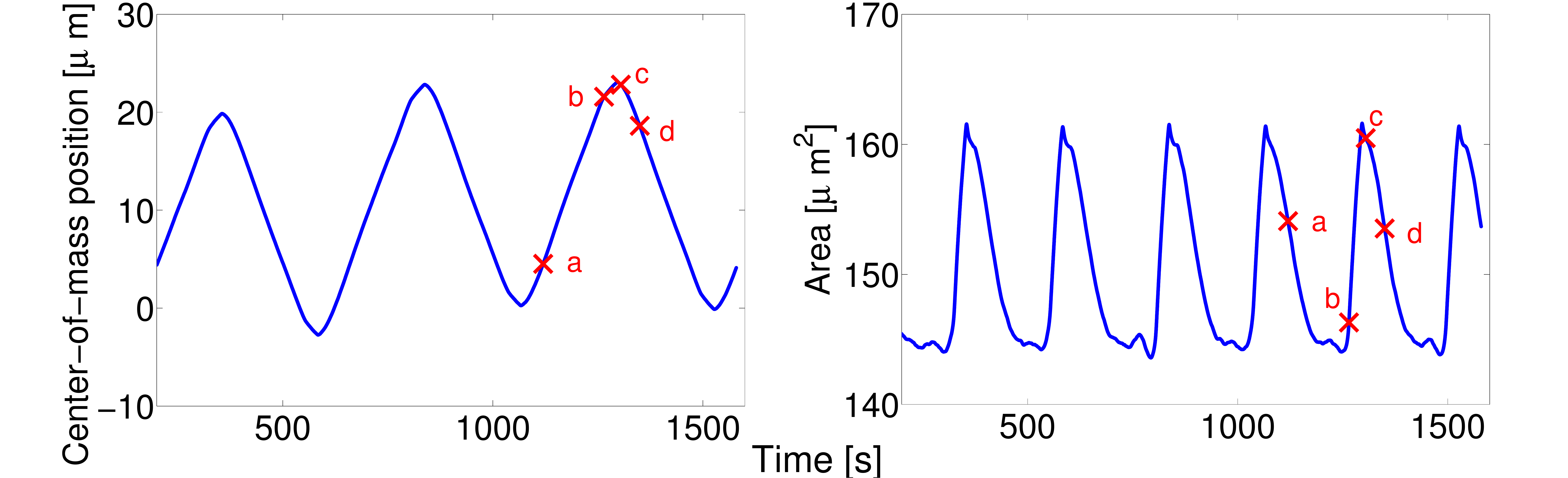}
\caption{TOP: Cell shape ($\phi$), actin promoter ($\rho_a
\phi$), and myosin ($\rho_m \phi$)
distribution during a reversal event in periodic migration.  Color plots are rescaled by $1$,$1.4 \um^{-2}$, and $0.55 \um^{-2}$ respectively.  Cell
velocity is indicated by an arrow.  Total width of stripe is $w = 6$
\um; (dashed lines).  BOTTOM: Center-of-mass
position ($\bar{y} = \frac{1}{A} \int d^2r\, y \phi(x,y)$) and area ($A
= \int d^2r \, \phi$) of cell as a function of time.  Full parameters for
all simulations are listed in the Appendix.}
\label{fig:periodic}
\end{figure}

{\it Simulation of periodic migration.} Numerical evaluation of \eqs \ref{eq:phasefield}-\ref{eq:rd_myo}
shows spontaneous emergence of periodic motion. 
An initially circular cell contracts to the stripe,
polarizes, migrates one way, then reverses and migrates in the
other direction.  We present one reversal in \fig
\ref{fig:periodic}.  When the cell is polarized ({$\rho_a$} is segregated
on one side), the cell contracts while crawling in the direction of
its polarization (point a in \fig \ref{fig:periodic}).  As the cell
contracts, it depolarizes (b).  The unpolarized cell
expands quickly, but does not crawl significantly.  As the
cell grows, it suddenly re-polarizes (c) and begins to travel in the
direction opposite to its initial direction.  As the cell moves,
myosin localizes to the cell rear {\cite{myosin_note}}, and the cell begins to
contract again (d).  Each reversal corresponds to one peak
in cell area.

Several questions arise: 1) How does cell
polarization control the cell's growth and contraction? 2) Why
does the cell depolarize at small areas and repolarize at large ones?
3) Why does the cell repolarize in a direction {\it opposite} to its
original motion?  We address these questions by reducing our
model to a significantly simpler 1D one.  

{\it Reduction to 1D model.}  We neglect adhesions and advection of $\rho_a$.
The latter is not strictly justified, as the Peclet number $\textrm{Pe} =
V_{\textrm{cell}} \lc / D_a$ is of order unity ({\lc is the cell length and $V_{\textrm{cell}}$ its velocity}), but we reproduce the 
essential aspects of the two-dimensional simulation
without fluid flow.  In migrating cells, myosin accumulates at the
back while actin is enriched at the front {\cite{myosin_note}}. 
We model these myosin dynamics
phenomenologically by letting
myosin go to the cell rear (where $\rho_a$
is low) with time lag $\tau$.  The simplified model for $\rho_a$ and
$\rho_m$ is 
\begin{align}
\partial_t\left(\phi \rho_a\right) = \partial_y\left[\phi D_a \partial_y \rho_a\right] + \phi f(\rho_a,\rho_a^\textrm{cyt}) \label{eq:wp}\\
\partial_t \rho_m^{f,b} = -\tau^{-1} \left[ \rho_m^{f,b} - \left(\rhoac - \rho_a^{f,b}\right) \right]
\end{align}
where $\rho^{f,b}_{a,m} = \rho_{a,m}(y_{f,b})$ and $\rhoac$ is the
equilibrium myosin when $\rho_a$ is zero.  The cell ``front'' is defined by $y_f > y_b$.  The cell shape
is $\phi(y,t) = \frac{1}{2}\left[ \tanh \frac{3(y-y_b)}{\epsilon} -
\tanh \frac{3(y-y_f)}{\epsilon} \right]$.  
{Actin polymerization causes local protrusion; myosin contraction causes local contraction.  The simplest form for the normal velocity of the edge is thus $\vb{v}_{\textrm{edge}}\cdot \hat{\vb{n}} = \alpha \rho_a - \beta \rho_m$, i.e.}
\begin{align}
\partial_t y_{f,b} = \pm (\alpha \rho_a^{f,b} - \beta \rho_m^{f,b})  \label{eq:motion}
\end{align}
This result can be rigorously justified in some limits by solving the Stokes equation (\eq \ref{eq:stokes}) in the presence of a planar front.  If $\epsilon / \lh \ll 1$ (sharp interface limit) and $\lc \gg \lh $, where $\lh = \sqrt{2 \nu_0 / \xi}$, we find 
$\alpha = \frac{\eta_a^0}{4 \nu_0}$ and $\beta = \frac{\eta_m^0 \lh}{2 \nu_0}$ (Appendix).  

\begin{figure}[ht!]
\includegraphics[width=90mm]{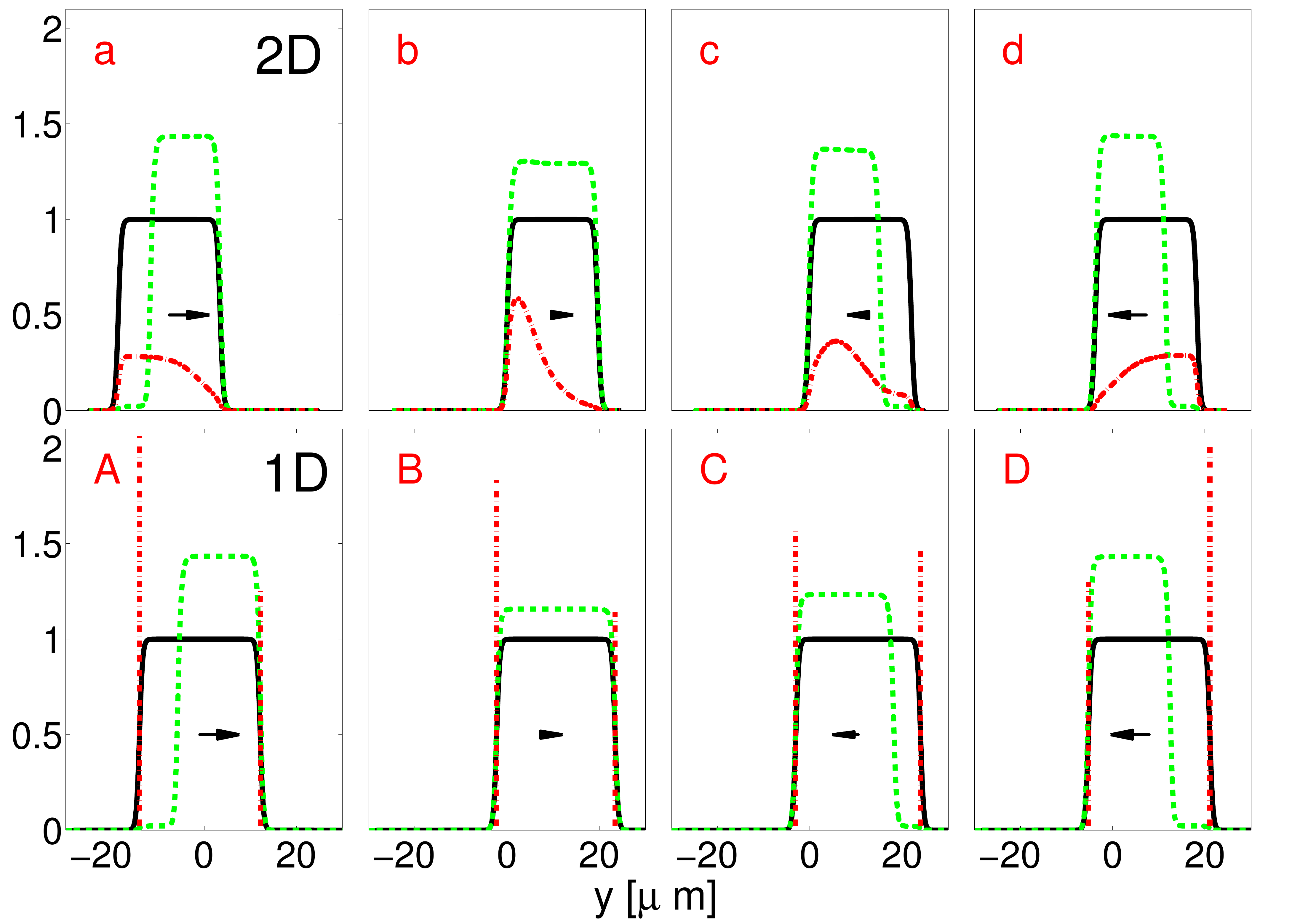}
\includegraphics[width=90mm]{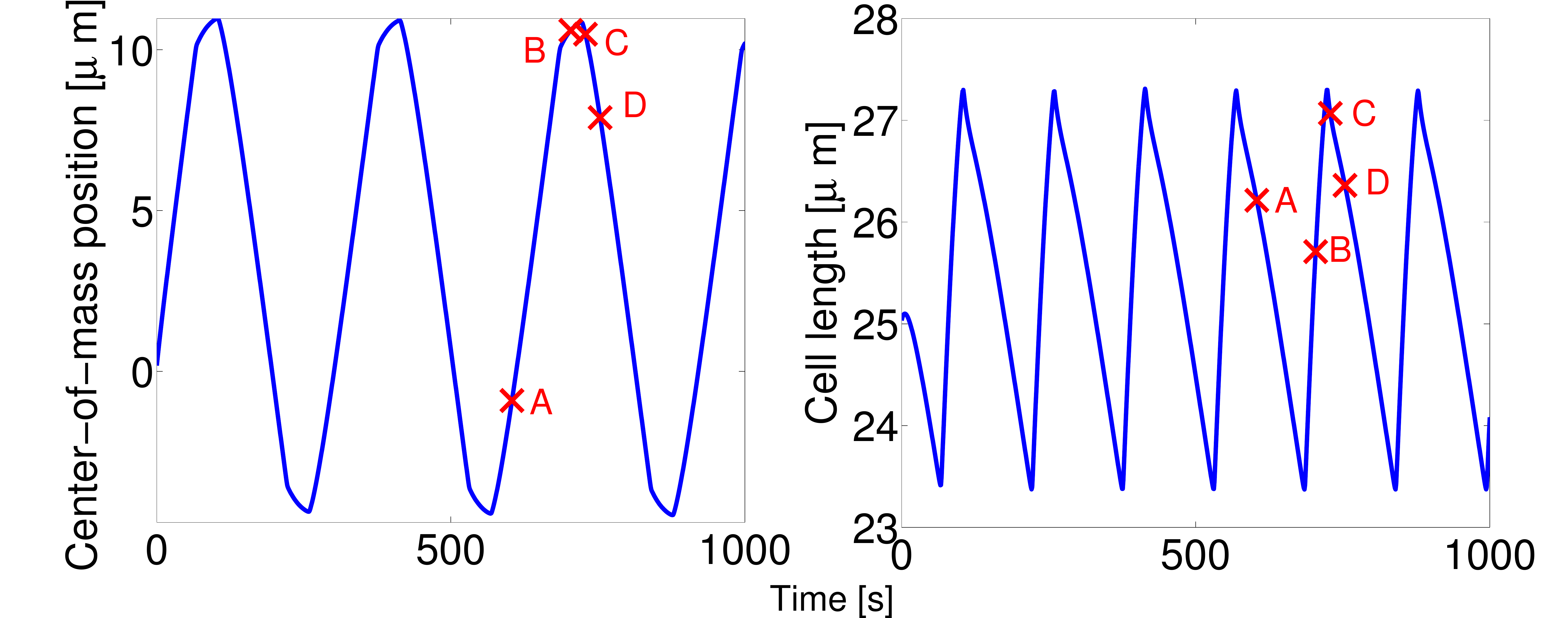}
\caption{Two- and one-dimensional models show highly similar behavior.  TOP: Centerline of \fig \ref{fig:periodic} a-d with $\phi$ (black solid line), $\phi \rho_a$ (green dashed line), and $\phi \rho_m$ (red dash-dotted line); axis is shifted for comparison to middle plot, MIDDLE: 1D model at comparable points in the periodic cycle (A-D).  
BOTTOM: Plot of position and size of periodically migrating 1D cell.}
\label{fig:oned}
\end{figure}

This limit is not necessarily applicable, as we have $\lh \approx 63 \um > \lc$.  Nevertheless, Eqs. \ref{eq:wp}-\ref{eq:motion} capture the essential features of periodic migration in \fig \ref{fig:periodic}.  We simulate them (\fig \ref{fig:oned}) and compare the 1D simulation to the centerline of \fig \ref{fig:periodic}.  

Our 1D model shows how the cell's shape changes and
polarization are coupled.  The cell shrinks if $\partial_t \lc =
\partial_t (y_f - y_b) = \alpha (\rho_a^f+\rho_a^b)-\beta
(\rho_m^f+\rho_m^b)$ is negative.  To find when this is true, we need
$\rho_a^{f,b}$.  {We use the analysis of Mori et al.
\cite{mori2008wave,mori2011asymptotic}, who proposed the wave-pinning reaction-diffusion model we apply in \eq \ref{eq:wp}.  
Their solutions
would be exact if $\epsilon\to0$ (sharp interface limit) and the cell were slow-moving, $\textrm{Pe} \ll 1$.}
$\textrm{Pe}$ is not small, but these solutions provide a valuable
qualitative guide to the cell's polarization as a function of its
size.  We use the simplified reaction kinetics
$\tilde{f}(\rho_a,\rho_a^{\textrm{cyt}}) \equiv -k \rho_a (\rho_a - h) (\rho_a - m
\rho_a^{\textrm{cyt}})$, which reproduce the phenomenology of the full
kinetics and permit analytical solutions.  {$h$ and $m$ are parameters related to the steady states of $\rho_a$ \cite{mori2008wave}.}  Mori et al. find two
homogeneous and linearly stable steady states, $\rho_a(y) = 0$ and
$\rho_a(y) = \frac{m N_a^{\textrm{tot}}}{\lc(1+m)}$, where
$N_a^{\textrm{tot}} = \int_{0}^L dy \, (\rho_a +
\rho_a^{\textrm{cyt}})$ is the conserved total number of {actin promoter molecules in either membrane-bound or cytosolic form}  ($N_a^{\textrm{tot}} = \int d^d r (\rho_a + \rho_a^{\textrm{cyt}})
\phi$ in the phase field model).  
Ref. \cite{mori2008wave} also finds a polarized state with a stationary front connecting a
region with $\rho_a = 2h$ to $\rho_a = 0$; the length of the region
with large $\rho_a$ is $y_p = \frac{N_a^{\textrm{tot}}}{2 h} -
\frac{\lc}{m}$.  The cell can only polarize if $y_p < \lc$, i.e. $\lc
> L_{\textrm{depol}} {\equiv} \frac{m N_a^{\textrm{tot}}}{2 h (m+1)}$.  This 
causes the cell to depolarize at small lengths, partially
answering Question 2 above.

Why doesn't the cell immediately repolarize when $\lc >
L_{\textrm{depol}}$?  The
homogeneous state $\rho_a(y) = \frac{m N_a^{\textrm{tot}}}{\lc(1+m)}$
is linearly stable; even though the cell can support a
polarized state if $\lc > L_{\textrm{depol}}$, it will not reach that
state without a perturbation beyond a certain threshold.  Numerically
evaluating \eq \ref{eq:wp} in a cell of fixed size, we find that this
threshold decreases with increasing cell size; larger cells are easier
to polarize.  (For the full kinetics $f(\rho_a,\rho_a^\textrm{cyt})$, this threshold can decrease to zero \cite{mori2011asymptotic}.)  Others \cite{meyers2006potential,holmes2012modelling,maree2012cells} have also suggested that cell shape influences signaling, polarization, and response to stimuli.

What perturbation causes the cell's repolarization?  Within the
1D model the only possibility is the moving edge. {If a cell edge expands faster than $\rho_a$ can be transported by diffusion or converted from cytosolic form, $\rho_a$ will be depleted near the expanding edge.  Explicitly: if we numerically solve \eq \ref{eq:wp} for an initially homogeneous cell with one edge expanding, the cell always polarizes to a state with low $\rho_a$ near the expanding edge. }  
Depletion sets
the direction in which the repolarization occurs.  As the cell
expands, both edges have high $\rho_a$ but one has lower $\rho_m$
(Figs. \ref{fig:periodic},\ref{fig:oned}).  {Edge normal velocity is set by \eq \ref{eq:motion}: actin polymerization causes expansion, but local contraction from myosin decreases the edge's velocity.  Therefore, the edge with low $\rho_m$ expands faster, leading to more depletion of
$\rho_a$ near that edge.}  When this depletion crosses the threshold of
patterning, a polarized state forms with low $\rho_a$ near the
quickly-moving edge, and high $\rho_a$ near the slowly-moving edge: the cell polarizes in the direction of
higher myosin.  Myosin keeps the memory of the cell's direction: if
it becomes uniform before the cell repolarizes, this information
is lost.  

We have now answered our questions: 1) Cell shape is
set by $\rho_a$ and $\rho_m$ via \eq
\ref{eq:motion}, and this is controlled by the cell
polarization.  2) At small cell sizes, \eq \ref{eq:wp} does not
support a polarized state, but as the cell expands, the polarized
state and homogeneous state are both stable.  Polarization requires a
perturbation to $\rho_a$ larger than a threshold, which decreases as the cell grows.  3) Repolarization is initiated
by depletion of $\rho_a$ near an expanding cell boundary; myosin
makes the previous ``back'' of the cell expand more slowly,
ensuring the cell polarizes in a direction opposite to its previous movement.  

We calculate the amplitude of periodic migration analytically by
using the results of \cite{mori2008wave} and making some additional assumptions.  We assume the cell
depolarizes at length $L_{\textrm{depol}}$ as above and
repolarizes in the direction of high myosin at a
critical length $L^*$.  
The value of $L^*$ would depend on the details of the cell's
motion, the diffusion coefficient $D_a$, and the threshold for
perturbations.  
We expect that the dominant contribution to the cell's displacement
over time will be the distance that it crawls while polarized; when the cell is polarized, it contracts.  We can then approximate the amplitude of periodic migration
as $A = v_{\textrm{cm}} t_{\textrm{contract}}$ where $v_{\textrm{cm}}$
is the cell center of mass velocity in the contraction phase,
and $t_{\textrm{contract}}$ the time required to contract from $L^*$
to $L_{\textrm{depol}}$.  Using \eq \ref{eq:motion} and the results of
\cite{mori2008wave}, we find that in the polarized state, $\partial_t \lc
\approx 2 h \alpha - \beta(2\rhoac - 2h)$ (assuming the myosin is at its equilibrium value $\rho_m = \rhoac-\rho_a$).
Similarly, $v_\textrm{cm} \approx h(\alpha + \beta)$.  We find
\begin{equation}
A = \left(\frac{L^*-L_{\textrm{depol}}}{2}\right)\frac{\gamma+1}{\gamma_c-\gamma} \label{eq:amplitude}
\end{equation}
where $\gamma = \alpha/\beta$ and $\gamma_c = (\rhoac-h)/h$.  For the cell to contract while polarized, $\gamma < \gamma_c$.  {The cell must also grow while unpolarized for periodic migration to occur; this condition depends on $N_a^{\textrm{tot}}$.}

For the amplitude of periodic migration to become large, protrusion and retraction must be balanced so that $\gamma-\gamma_c$ is small.  However, this requirement can be weakened by the cell's internal dynamics, which we have mostly neglected in deriving \eq \ref{eq:amplitude}.  If we assume a large viscous resistance to changes in size, we suppress the rate of contraction and expansion by a factor $\lambda$, where $\lambda \ll 1$.  If the cell's contraction is slowed, but crawling is not, the amplitude of periodic migration increases significantly, as $t_{\textrm{contract}} \sim 1/\lambda$, and $A \approx t_{\textrm{contract}}v_{\textrm{cm}}$, so $A \sim 1/\lambda$ becomes large.  

{\it Additional emergent behaviors}.  Depending on initial conditions and micropattern width, other behaviors are observed.  These include steady crawling, turning, and bipedal motion (see Appendix).  The bipedal motion resembles that seen theoretically and experimentally by Barnhart et al. \cite{barnhart2010bipedal}.  Turning has been studied by Rubinstein et al. \cite{mogilner2010actin,rubinstein2005multiscale}; see also \cite{tjhung2012spontaneous}.  We plan to address the origin of these effects within our model in future work.  

If periodic migration in \cite{fraley2012dimensional} arose
through precisely the mechanism we have described, the cell area would oscillate with a period half that of the
cell's migration {and myosin reorientation would lag the reversal of cell direction} (Figs. \ref{fig:periodic}-\ref{fig:oned}).  It would 
be interesting to experimentally quantify total
surface area {and myosin localization} of periodically migrating cells.
We present this study primarily as an example of complex behaviors that develop when cell polarization is coupled
to cell shape.  However, our mechanism of periodic
migration may be more general if cell polarization is
coupled to other mechanical properties.  
Cell-surface adhesion is a natural choice, as periodic motion arises in \cite{fraley2012dimensional} when the 
adhesion protein zyxin is depleted.
If cells only polarize when sufficiently adherent to the surface, and
this adhesion changes with cell motion, our periodic migration scheme
may be recapitulated with adhesion in place of cell area.

Periodic migration as observed in our simulations is a new,
interesting, and tractable example of the complex dynamics resulting
from coupling cell shape and polarity.  {Periodic migration requires a balance between contraction and protrusion  (\eq \ref{eq:amplitude}), but its existence is robust to many model details.  Within our larger model, individual adhesions can be neglected, as can the $\rho_a$-dependence of $D_m$.  In the 1D model, we have ignored hydrodynamics entirely.  Removing features or varying parameters (Appendix) changes migration amplitude, but if the fundamental aspects illustrated by the 1D model are present, periodic migration exists.}
Therefore, we believe periodic migration
could be observed in other models of eukaryotic cell motility
that couple polarity and cell shape
\cite{holmes2012comparison,wolgemuth2011redundant,ziebert2012model,shao2010computational,maree2012cells}, especially those
using the wave-pinning polarity mechanism \cite{mori2008wave}.
Randomly-occurring reversals 
without periodicity have been observed by
Ziebert and Aronson \cite{ziebertpreprint}; their model may only lack
a memory.  Our one-dimensional model suggests the essential elements
required for periodic migration, and emphasizes the role of myosin in preserving the memory
of the cell's initial direction.  Our model for cells on adhesive
micropatterns and the analytical tools we developed to study periodic
migration may be useful in understanding more complex behavior on
micropatterns, including ``dimension sensing''
\cite{chang2013guidance}, response of fibroblasts to cross-hatched
patterns \cite{doyleyamada}, and polarization in response to
asymmetric micropatterns
\cite{thery_review,thery2006anisotropy,mahmud2009directing}.  In all
of these cases, cell polarity is coupled to the underlying
micropattern.  The coupling of micropattern shape, cell shape, and
cell polarization studied here will be essential to a deeper
understanding of these problems.  

This work was supported by NIH Grants P01 GM078586 and R01 GM096188, NSF Grant DMS 1309542, and by the Center for Theoretical Biological Physics.

\newpage
\onecolumngrid
\appendix
\section{Turning and bipedal motion}

\begin{figure}[ht!]
\begin{center}
\includegraphics[width=85mm]{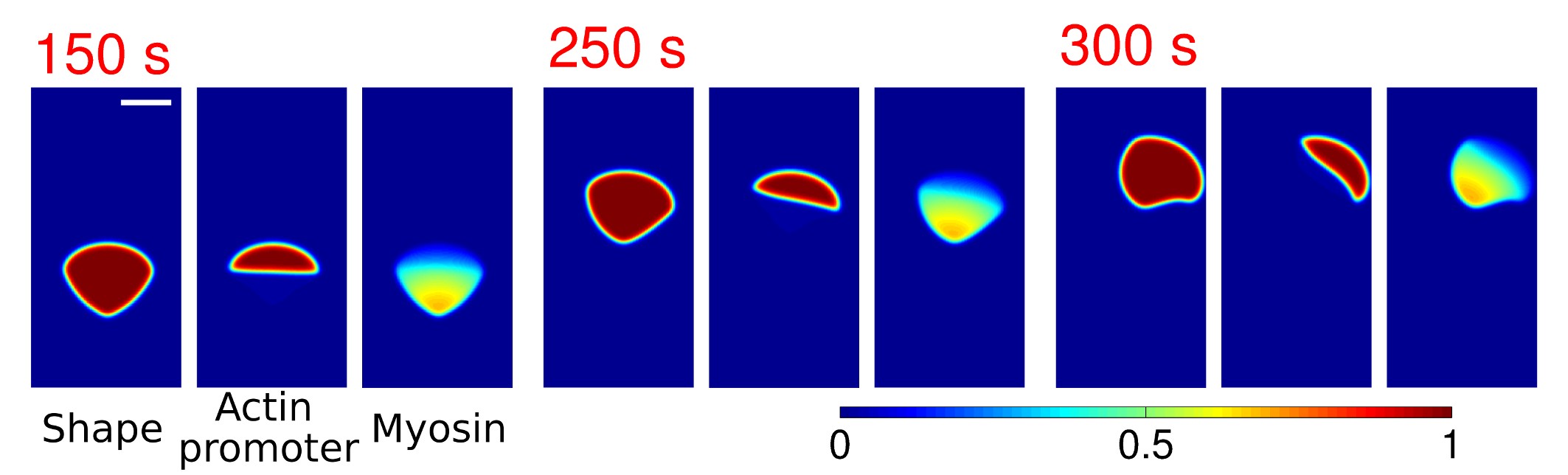}\\
\includegraphics[width=85mm]{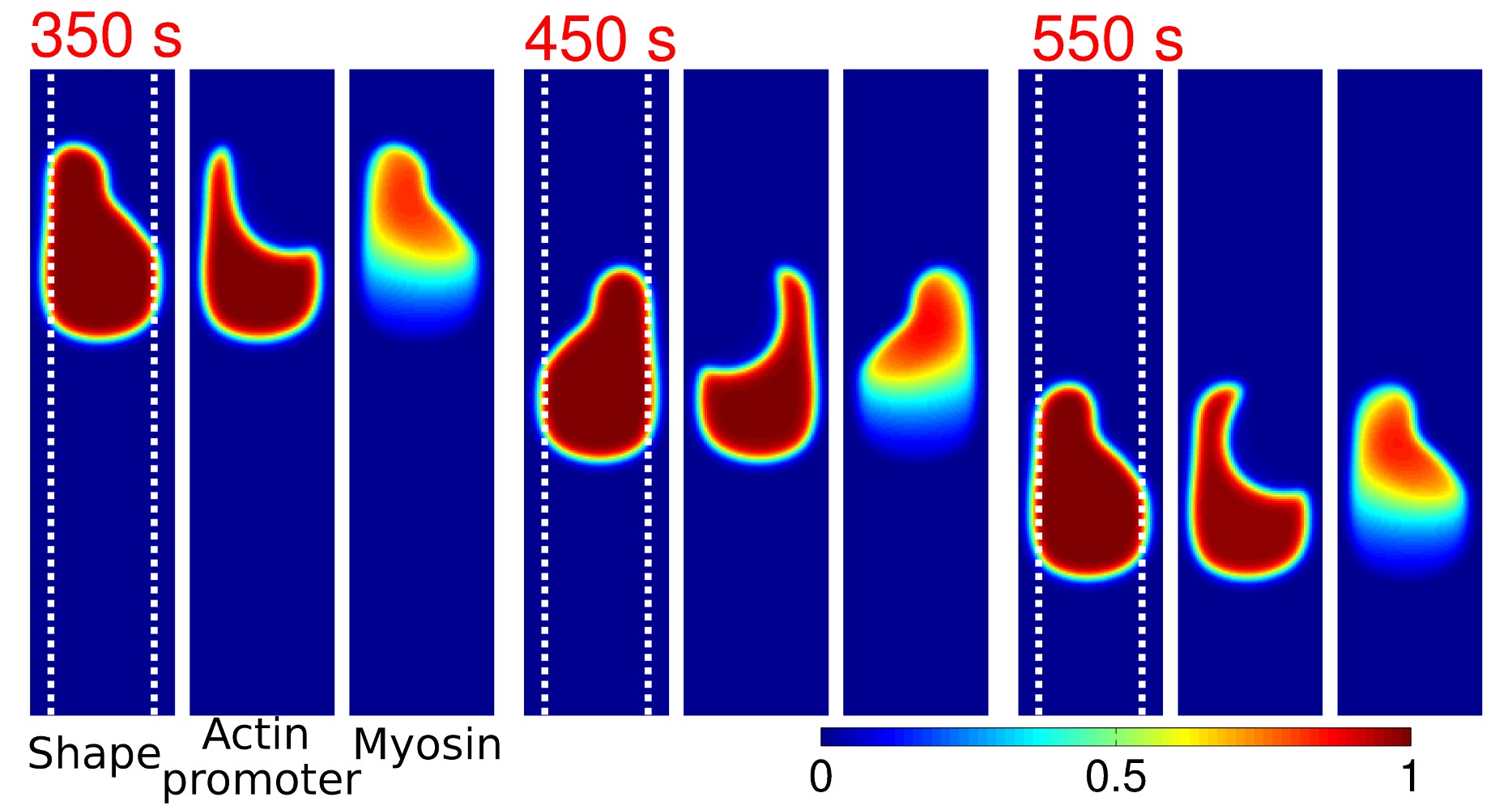}
\end{center}
\caption{Many types of cell crawling appear, including bipedal motion and turning.  Cell shape (phase field $\phi$), actin promoter ($\rho_a \phi$), and myosin ($\rho_m \phi$) of cells at different times.  Color plots are rescaled by $1$, $1.4$ $\um^{-2}$, and $0.8$ $\um^{-2}$, respectively.  TOP: Cell turning; total width of stripe is $w = 40$ \um \, (not in image; scale bar indicates 10 \um).  BOTTOM: Bipedal motion.  Total width of stripe is $w = 10$ \um (dashed lines).}   
\label{fig:other}
\end{figure}

\section{Sharp interface derivation}

We will derive the sharp interface results presented in the text.  These are that the front and back interface velocities are given by $\pm \alpha \rho_a^{f,b}\mp\beta\rho_m^{f,b}$ with 
\begin{equation}
\alpha = \frac{\eta_a^0}{4 \nu_0}, \; \; \; \beta = \frac{\eta_m^0 \lh}{2 \nu_0}, \label{eq:alphabeta}
\end{equation}
where $\lh^2 = 2\nu_0/\xi$.  In order to get these results, we will assume the sharp interface limit $\epsilon / \lh \ll 1$, and also that the cell's size $\lc$ is much larger than $\lh$.  We will also assume that the interface's curvature is not relevant.  

Our Stokes equation for the cell's cytoskeletal velocity $\vb{u}$ is
\begin{equation}
\nabla \cdot \left[ \nu(\phi) \left( \nabla \vb{u} + \nabla \vb{u}^T \right) \right]  + \nabla\cdot\sigma_{\textrm{myo}} + \nabla \cdot\sigma_{\textrm{poly}} + \vb{F}_{\textrm{mem}} +\vb{F}_{\textrm{adh}} - \xi \vb{u} = 0
\end{equation}
where $\nu(\phi) = \nu_0 \phi(\rb)$ and the active stresses are given by
\begin{align}
\sigma_{\textrm{myo}} &= \eta_m^0 \phi \rho_m \vb{I} \\
\sigma_{\textrm{poly}} &= -\eta_a^0 \phi \rho_a \delta_\epsilon \hat{\vb{n}}\hat{\vb{n}}
\end{align}
where $\vb{I}$ is the identity tensor, $\delta_\epsilon = \epsilon (\nabla \phi)^2$, and $\hat{\vb{n}} $ is the unit normal vector to the cell boundary.  $\vb{F}_{\textrm{adh}}$ contains stochastic adhesion forces, which we ignore.  We note that these adhesion forces may in some limits only renormalize $\xi$ \cite{walcott2010mechanical}, so it may be appropriate to think of the $\xi$ as an effective value larger than that given in the simulation.     The membrane forces are derived from a phase field approximation to the Helfrich energy and surface tension (see, e.g. \cite{du2009energetic,shao2012coupling}), $\vb{F}_{\textrm{mem}} = \vb{F}_{\textrm{tension}} + \vb{F}_{\textrm{bend}}$ with
\begin{align}
\vb{F}_{\textrm{tension}} &= -\gamma \left( \epsilon \nabla^2 \phi - \frac{G'}{\epsilon}\right) \nabla \phi \\
\vb{F}_{\textrm{bend}} &= \kappa \epsilon \left( \nabla^2 - \frac{G''}{\epsilon^2}\right) \left( \nabla^2 \phi - \frac{G'}{\epsilon^2}\right) \nabla \phi
\end{align}
where $G(\phi) = 18 \phi^2 (1-\phi)^2$ and $G'$ and $G''$ denote derivatives of $G$ with respect to $\phi$.  

We are interested in creating an effectively one-dimensional model.  We approximate our cell's complex shape by an effectively one-dimensional front that minimizes the tension and bending energies (i.e.  $\vb{F}_{\textrm{tension}} = \vb{F}_{\textrm{bend}} = 0$).  This will be true if $ \epsilon \nabla^2 \phi = \frac{G'}{\epsilon}$, or (for a front in the y direction)
\begin{equation}
\phi_I(y) = \frac{1}{2} \left( 1 + \tanh(3 y / \epsilon) \right)
\end{equation}

Under this assumption, the Stokes equation becomes

\begin{equation}
\tilde{\nu} \partial_y \left[\phi_I(y) \partial_y u \right] + F_{\textrm{myo}}\left[\phi_I\right] + F_{\textrm{poly}}\left[ \phi_I \right] -\xi u = 0 \label{eq:stokesapp}
\end{equation}
where $\tilde{\nu} = 2 \nu_0$, $F_{\textrm{poly}}\equiv\partial_y \sigma_{\textrm{poly}}$ and $F_{\textrm{myo}}\equiv\partial_y \sigma_{\textrm{myo}}$  We will look at the two force terms separately, since this equation is linear and we can superimpose the two resulting velocity fields.  
We will also assume that the densities $\rho_a$ and $\rho_m$ do not vary quickly at the front, so that $F_{\textrm{poly}} \approx -\eta_a^0 \rho_a \partial_y (\phi_I \de)$ and $F_{\textrm{myo}} \approx \eta_m^0 \rho_m \partial_y \phi_I$.  We illustrate the resulting field $\phi_I$ and the forces in \fig \ref{fig:forces} below.

\begin{figure}[ht!]
\begin{centering}
\includegraphics[width=100mm]{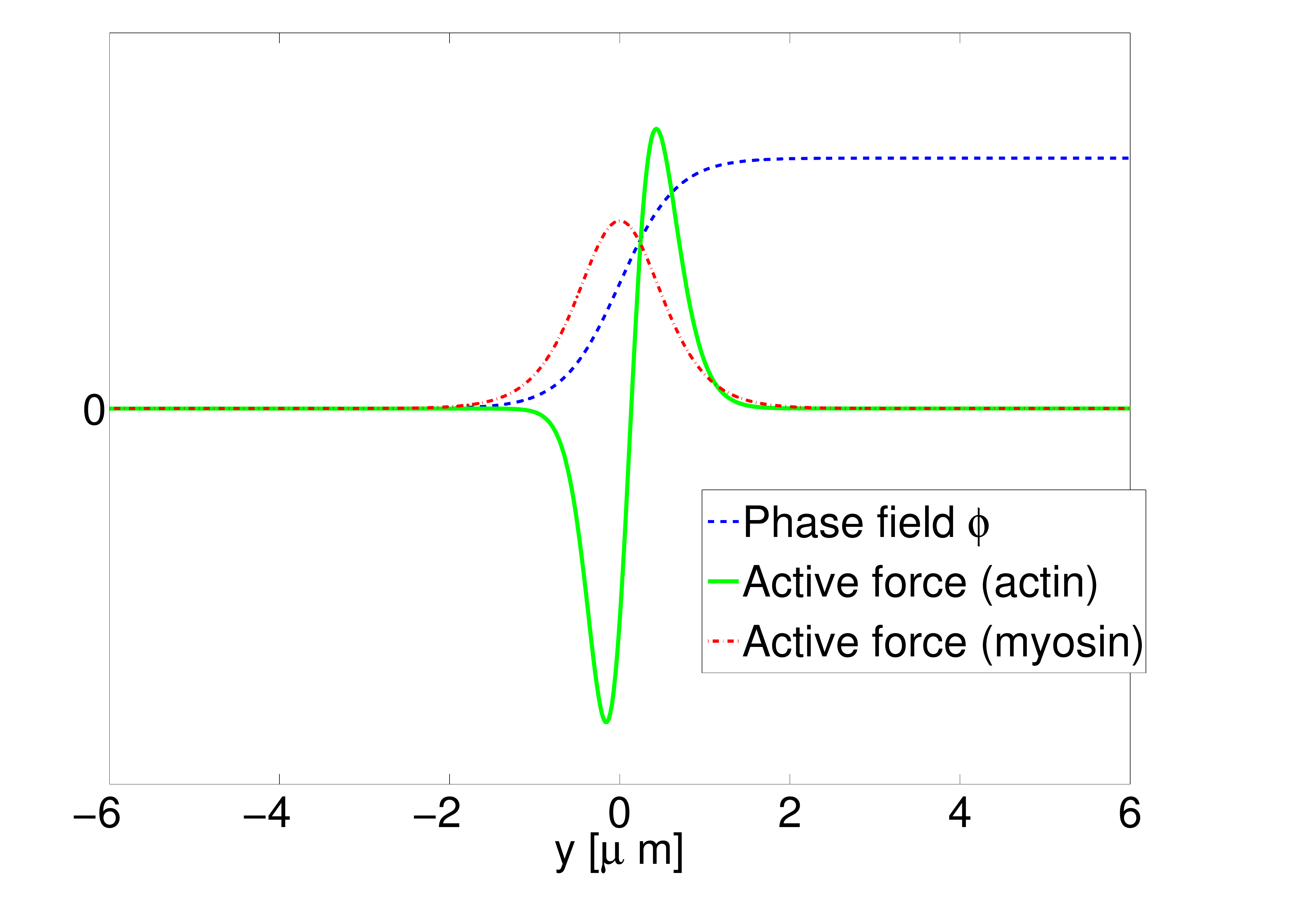}
\caption{We show the phase field and active forces at the interface at  $y=0$.  Here the phase field $\phi_I  = \frac{1}{2} \left( 1 + \tanh(3 y / \epsilon) \right)$.  The active force due to actin polymerization is $F_{\textrm{poly}} \approx -\eta_a^0 \rho_a \partial_y (\phi_I \de)$ and the active force due to myosin contractility is $F_{\textrm{myo}} \approx \eta_m^0 \rho_m \partial_y \phi_I$.  $\epsilon = 2 \um$ in this figure.}
\label{fig:forces}
\end{centering}
\end{figure}

\subsection{Active force due to actin polymerization}
We start by rewriting \eq \ref{eq:stokesapp} with $F_{\textrm{poly}} = -\eta_a^0 \rho_a \partial_y (\phi_I \de)$ and $F_{\textrm{myo}} = 0$.  (We will use linearity to rescue the complete result later.)  Rescaling our lengths to $r = y / (\lh)$ where $\lh^2 = \tilde{\nu}/\xi$ and defining $\delta = \epsilon / 3 \lh$, we find
\begin{equation}
\partial_r \left[\left\{1+\tanh (r/\delta)\right\} \partial_r u \right] - \frac{\chi}{\delta} \partial_r\left[\left\{1+\tanh(r/\delta) \right\} \, \sech^4(r/\delta)\right] - 2 u = 0
\end{equation}
where $\chi = \frac{3}{4} \eta_a^0 \rho_a / \tilde{\nu}$.
We can't solve this equation exactly, but can develop an asymptotic approximation in the sharp interface limit of $\epsilon \ll \lh$ ($\delta \ll 1$).  In particular, we can see that in the sharp interface limit, the term $\sech^4 r/\delta$ can be neglected everywhere but near the front position, $r = O(\delta)$.  Moving to the stretched variable $z = r/\delta$, and defining $U(z) = u(r)$ for convenience, 
\begin{equation}
\partial_z \left[(1+\tanh z) \partial_z U \right] - \chi \partial_z\left[(1+\tanh z) \, \sech^4 z\right] - 2 \delta^2 U = 0
\end{equation}
To $O(\delta^0)$, we can neglect the last term on the right.  The remaining ODE can be directly integrated:
\begin{equation}
U(z) = A\left(z - \frac{1}{2} e^{-2z} \right) + B + \chi \left\{ \frac{4}{(1+e^{-2z})^2} - \frac{8}{3 (1+e^{-2z})^3} \right\}
\end{equation}
We cannot consistently apply the boundary conditions $u(r\to\pm\infty) = 0$ to this solution; we need to match it to the solution in the outer region.  However, the outer regions to the right and left of the front have two distinctly different characters.  For $r \gg \delta$, $1+\tanh(r/\delta) \approx 2$, and the outer expansion is
\begin{equation}
\partial_r^2 u_R - u_R = 0
\end{equation}
and we can immediately determine $u_R = C e^{-r}$, neglecting the solution that diverges as $r\to\infty$.  Matching to the interior solution yields the requirement $C = B + \frac{4}{3} \chi$ and $A = -\delta C$.
However, for $r \ll -\delta$, $1+\tanh(r/\delta)$ approaches zero; $\delta$ is a singular perturbation to the outer equation in the left region.  For $r \ll -\delta$, $1+\tanh(r/\delta) \approx 2 e^{2 r/\delta}$, and so
\begin{equation}
\partial_r(e^{2 r/\delta} \partial_r u_L) - u_L = 0
\end{equation}
which can be solved to find 
\begin{equation}
u_L = D e^{-r/\delta} K_1(\delta e^{-r/\delta})
\end{equation}
where $K_1$ is the modified Bessel function of order 1, and we have dropped the solution that diverges as $r\to-\infty$.  To match this to the interior solution, we choose $r = \delta z_c$ with $z_c$ fixed but large (and negative), and look at the behavior as $\delta \to 0$: 
\begin{align}
u_L &\sim D \left[ \frac{1}{\delta} + \frac{e^{-2z_c}}{2} \delta \left\{\ln\delta-z_c-\ln 2+ \gamma_E -1/2\right\} \right]\\
    &\sim D \left[ \frac{1}{\delta} + \frac{e^{-2z_c}}{2} \delta \ln \delta \right]
\end{align}  
where $\gamma_E$ is the Euler gamma, $\gamma_E = 0.5772\dots$.  
We match to the interior solution at $z = z_c$ with $z_c \ll -1$, 
\begin{equation}
U \sim -\frac{1}{2} A e^{-2z_c} + B
\end{equation}

Matching then requires that $-A = D \delta \ln \delta$ and $B = D/\delta$.  Combining this with our earlier matching requirements, $C = B + \frac{4}{3} \chi$ and $A = -\delta C$, we find:
\begin{align}
A/\chi &= \frac{4}{3} \frac{\delta^2 \ln \delta}{1-\delta\ln\delta}\\
B/\chi &= -\frac{4}{3} \frac{1}{1-\delta\ln\delta}\\
C/\chi &= -\frac{4}{3} \frac{\delta \ln \delta}{1-\delta\ln\delta}\\
D/\chi &= -\frac{4}{3} \frac{\delta}{1-\delta\ln\delta}
\end{align}
Importantly, because $C \to 0$ as $\delta \to 0$, in the sharp interface limit, there is no long-range velocity induced by the actin promoter at the interface.  

\begin{figure}[h]
\centering
\includegraphics[width=100mm]{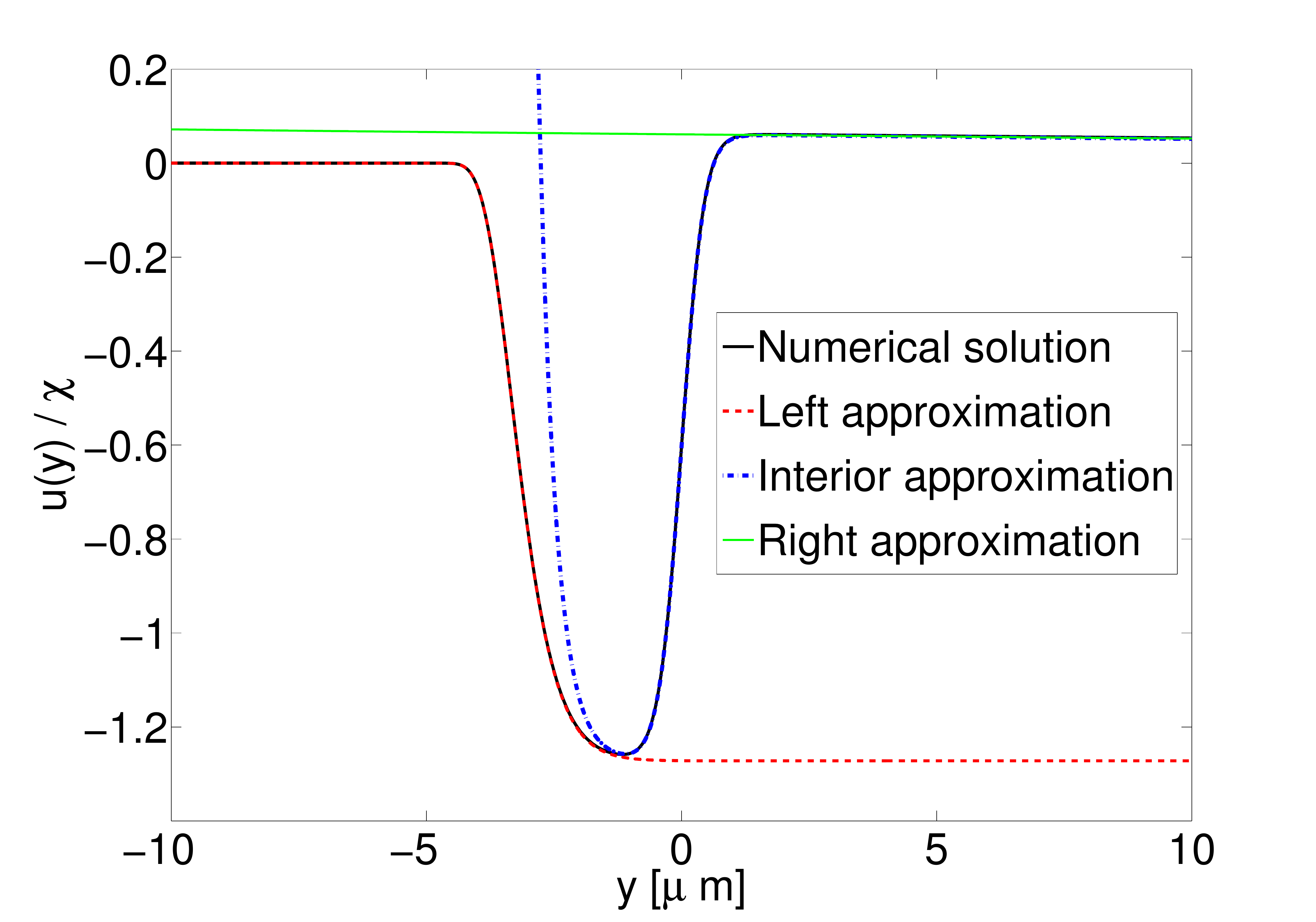}
\caption{Velocity of fluid due to the presence of actin promoter at the cell boundary.  Interface is at $y=0$ as above, i.e. $\phi = \frac{1}{2}\left[1+\tanh(3y/\epsilon)\right]$.  Here $\epsilon = 2$ \um, $\lh = 63$  \um, i.e. $\delta \approx 0.01$.}
\label{fig:actin}
\end{figure}

Our asymptotics provide an excellent approximation to the full numerical solution (\fig \ref{fig:actin}).  It also allows us to determine the interface velocity, $u(0)$ (using the interior solution).  We find in the sharp interface limit that
\begin{equation}
u_{\textrm{interface}} = -\frac{2}{3} \chi \equiv -\alpha \rho_a
\end{equation}
where
\begin{equation}
\alpha = \frac{\eta_a^0}{2 \tilde{\nu}} = \frac{\eta_a^0}{4 \nu_0}.
\end{equation}
This is the result given in the main paper.  It is only the leading order term; higher-order terms that depend on $\delta$ can also be obtained from the solution above.  The process for myosin is very similar, but we will find that a long-range (on the order of $\lh$) velocity will be induced, unlike the actin promoter case.

\subsection{Myosin force}
We start by rewriting \eq \ref{eq:stokesapp} with  $F_{\textrm{myo}} = \eta_m^0 \rho_m \partial_y \phi$ and $F_{\textrm{poly}} = 0$.  Rescaling our lengths to $r = y / (\lh)$ where $\lh^2 = \tilde{\nu}/\xi$ and defining $\delta = \epsilon / 3 \lh$, we find
\begin{equation}
\partial_r \left[\left\{1+\tanh (r/\delta)\right\} \partial_r u \right] + \mu \partial_r\left[1+\tanh(r/\delta)\right] - 2 u = 0
\end{equation}
where $\mu = \eta_m^0 \rho_m \lh / \tilde{\nu}$.  Note that unlike $\chi$ in the actin promoter case, $\mu$ does have an explicit dependence on the hydrodynamic length scale \lh.

We develop an asymptotic approximation in the sharp interface limit of $\epsilon \ll \lh$ ($\delta \ll 1$).  Moving to the stretched variable $z = r/\delta$, and defining $U(z) = u(r)$,
\begin{equation}
\partial_z \left[(1+\tanh z) \partial_z U \right] + \mu \delta \partial_z\left[1+\tanh z \right] - 2 \delta^2 U = 0
\end{equation}
To linear order in $\delta$, the last term can be dropped, and the remaining equation can be easily integrated to find
\begin{equation}
U(z) = A\left(z - \frac{1}{2} e^{-2z}\right) + B - \frac{\mu \delta}{2} e^{-2z}
\end{equation}
The outer limits are the same as in the actin promoter case.  We then get the matching conditions $D/\delta = B, A+\mu\delta = -D \delta \ln \delta$, $B = C$, and $A = -\delta C$.  These can be solved to find 
\begin{align}
A/\mu &= -\frac{\delta}{1-\delta\ln\delta}\\
B/\mu &= \frac{1}{1-\delta\ln\delta}\\
C/\mu &= \frac{1}{1-\delta\ln\delta}\\
D/\mu &= \frac{\delta}{1-\delta\ln\delta}
\end{align}

\begin{figure}[h]
\centering
\includegraphics[width=100mm]{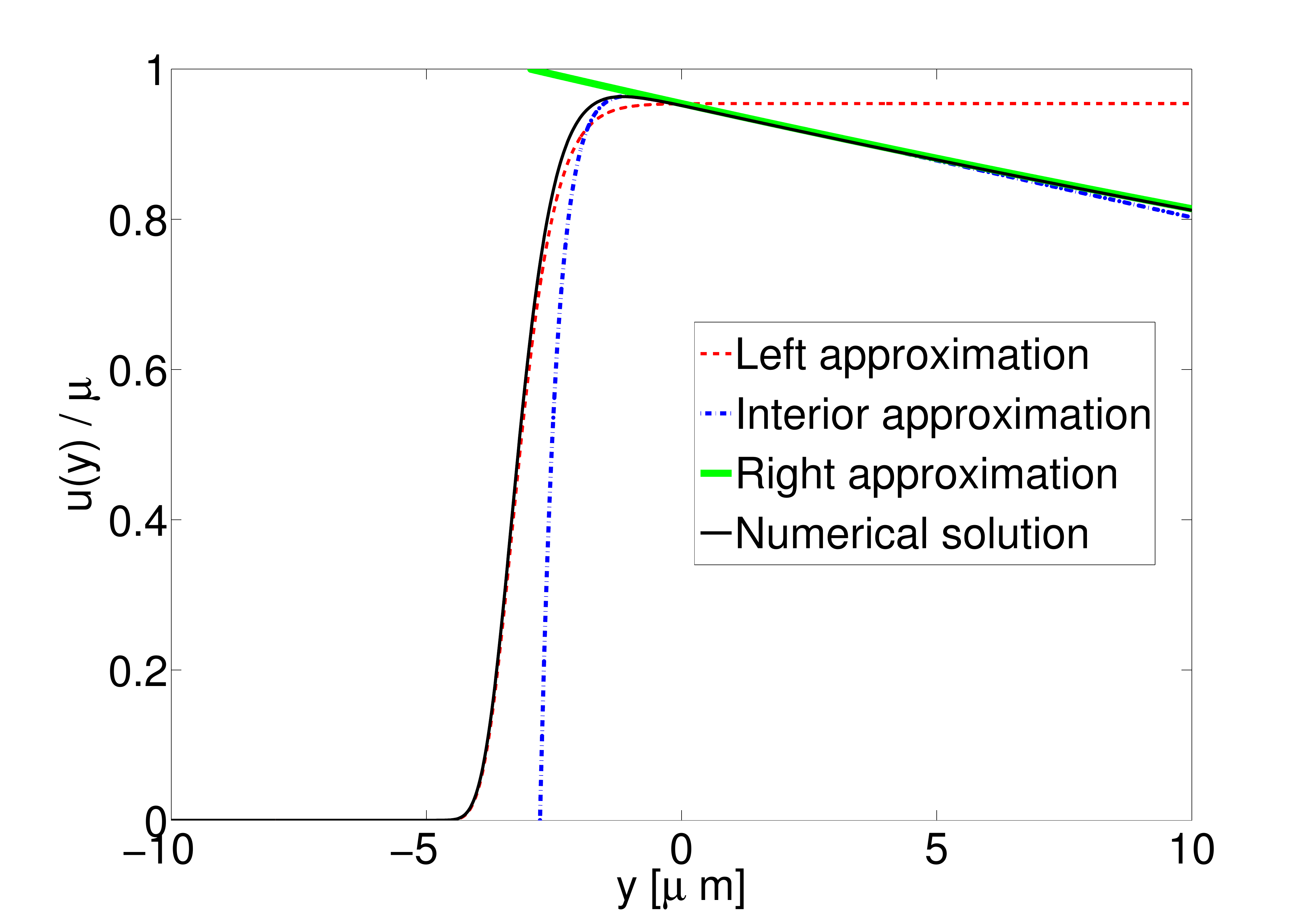}
\caption{Velocity of fluid due to the presence of myosin at the cell boundary.  Interface is at $y=0$ as above, i.e. $\phi = \frac{1}{2}\left[1+\tanh(3y/\epsilon)\right]$.  Here $\epsilon = 2$ \um, $\lh = 63$  \um, i.e. $\delta \approx 0.01$.  Note that even though $\delta$ is small, the velocity to the right of the interface is not small since $C$ is  $O(\delta^0)$.}
\label{fig:myosin}
\end{figure}

Note that $C$ does not vanish in the sharp interface limit: the presence of myosin at the interface leads to a velocity far away from the interface, $u(r) \approx \mu e^{-r}$.  Our asymptotic approximations are again an excellent approximation to the full numerical solution (\fig \ref{fig:myosin}).  

In the sharp interface limit, $u(0)$ becomes
\begin{equation}
u_{\textrm{interface}} = \mu \equiv \beta \rho_m
\end{equation}
where
\begin{equation}
\beta = \frac{\eta_m^0 \lh}{\tilde{\nu}} = \frac{\eta_m^0 \lh}{2 \nu_0}
\end{equation}

\subsection{When can we apply the sharp interface result?}

We argue that in the limit $L_{\textrm{cell}} \gg \lh$, we can neglect correlations between the cell edges.  We have been attempting to determine the velocity of the cell interface using only the actin promoter and myosin densities at the interface, but no information about the actin promoter and myosin throughout the cell, or the other interface of the cell.  When is this appropriate?  We have seen above that myosin at the cell interface induces a velocity in the cell body with a dependence of position of $e^{-y/\lh}$; if $L_{\textrm{cell}} \gg \lh$, one interface will not affect the other.  We have also neglected forces coming from {\it internal} gradients of the myosin-induced stress; once again, the characteristic length scale for these forces is $\lh$, and so they should not affect the velocity of the interfaces if $L_{\textrm{cell}} \gg \lh$.  The sharp interface results could be generalized to include all of these effects, but they produce additional complications, such as the need to track the details of myosin within the cell.  

\section{Tables of parameters used}

\subsection{Parameters used for all two-dimensional phase field simulations}
We mark with an asterisk the parameters that have been changed from the simulations presented in Ref. \cite{shao2012coupling}.  Parameters were originally chosen in \cite{shao2012coupling} to ensure that the cell velocity, actin flow velocity, and midline stress were close to experimentally reported values for keratocytes; in general, we have attempted not to change these values.  Where possible, we have given literature justification for these parameters.  

\subsubsection{Phase field and cell boundary properties}
\begin{tabular}{lccp{5cm}}
Parameter & Description & Value & Justification\\
\hline
$\gamma$ & Cell tension coefficient &  20 pN & Order-of-magnitude set in \cite{shao2010computational}\\
$\kappa$ & Cell bending coefficient & 20 pN $\um^2$ & Order-of-magnitude set in \cite{shao2010computational}\\
$\epsilon$ & Phase field width & 2 \um & Chosen to ensure smooth variation in $\phi$ \\
$\Gamma$ & Phase field relaxation parameter & 0.4 \um/s & Set in \cite{shao2012coupling} \\
\end{tabular}

\subsubsection{Cytoskeletal flow parameters}
\begin{tabular}{lccp{5cm}}
Parameter & Description & Value & Justification\\
\hline
$\nu_0$ & Viscosity of cytoskeletal flow & $10^3$ pN s / \um & Set roughly by \cite{bausch1998local}; see also \cite{rubinstein2009actin} \\
$\eta_a^0$ & Protrusion coefficient & 560 pN \um$^2$ & Chosen to reproduce shapes and other features in \cite{shao2012coupling} \\
$\eta_m^0$ & Myosin contractility coefficient & 60-61 pN \um$^*$ & Similar to that of \cite{rubinstein2009actin}; tuned to increase periodic migration amplitude\\
$\xi$ & Substrate friction coefficient & 0.5 Pa s / \um & Value arising from cell sitting on layer of water with height 2 nm \cite{evans1988translational} \\
\end{tabular}

\subsubsection{Reaction-diffusion parameters}
In the same wave-pinning kinetics as \cite{shao2012coupling} for the reaction term in the actin promoter equation,
\begin{equation}
f(\rho_a,\rho_a^\textrm{cyt}) = k_b \left(\frac{\rho_a^2}{K_a^2+\rho_a^2}+k_a\right)\rho_a^\textrm{cyt} -k_c \rho_a
\end{equation}
where, by the conservation of total actin promoter, $\int d^2 r \left( \rho_a(\vb{r}) + \rho_a^{\textrm{cyt}} \right) \phi(\vb{r}) = N_a^{\textrm{tot}}$, or, assuming the cytosolic actin promoter is well-mixed (uniform),
\begin{equation}
\rho_a^\textrm{cyt} = \frac{N_a^{\textrm{tot}} - \int d^2 r \rho_a(\vb{r}) \phi (\vb{r})}{\int d^2 r \phi(\vb{r})}.  
\end{equation}
We note that this formula was written incorrectly in the Supplementary Material of Ref. \cite{shao2012coupling}.  

\begin{tabular}{lccp{5cm}}
Parameter & Description & Value & Justification\\
\hline
$k_a$ & Unitless base activation rate & 0.01 \footnotemark[1] & Order-of-magnitude from \cite{mori2008wave}\\
$k_b$ & Overall activation rate  & 10 s$^{-1}$  & Order-of-magnitude from \cite{mori2008wave} \\
$k_c$ & Deactivation rate & 10 s$^{-1}$ & Order-of-magnitude from \cite{mori2008wave} \\
$K_a$ & Positive feedback threshold for actin promoter concentration & 1 \um$^{-2}$ & Order-of-magnitude from \cite{mori2008wave} \\
$D_a$ & Actin promoter diffusion coefficient & 0.8 \um$^{2}$/s & Typical membrane-bound protein diffusion coefficient \cite{postma2004chemotaxis}\\
$D_m^0$ & Myosin diffusion coefficient at zero actin concentration & 2 \um$^{2}$/s & Chosen in \cite{shao2012coupling} \\
$K_D$ & Myosin diffusion threshold, $D_m = D_m^0 / (1 + \rho_a / K_D)$ & 0.5 \um$^{-2}$ \footnotemark[2] & Chosen in \cite{shao2012coupling} \\
$N_{a}^{\textrm{tot}}$ & Total amount of actin promoter & 485$^*$ \footnotemark[3] & Roughly rescaled by cell size from value chosen in \cite{shao2012coupling} proportional to cell area \\
$\rho_{m}^0$ & Initial density of myosin & 0.3 \um$^{-2}$ & Chosen such that myosin stress corresponds to that estimated in \cite{rubinstein2009actin} \\
\end{tabular}
\footnotetext[1]{The units for this parameter were listed incorrectly in the Supplemental Material of Ref. \cite{shao2012coupling}.}
\footnotetext[2]{This is the value used in Ref. \cite{shao2012coupling}, though it was listed incorrectly in the Supplemental Material of that work.} 
\footnotetext[3]{This parameter is denoted by $\rho_a^{\textrm{tot}}$ in Ref. \cite{shao2012coupling}.}

\subsubsection{Adhesion parameters}
\begin{tabular}{lccp{5cm}}
Parameter & Description & Value & Justification\\
\hline
$N_{\textrm{adh}}$    & Number of adhesions & 1000$^*$  \footnotemark[4] & Roughly rescaled from value chosen in \cite{shao2012coupling} proportional to cell area \\
$F_\textrm{grip}^0$ & Characteristic gripping stress for gripping-slipping rupture & 5 Pa & Chosen in \cite{shao2012coupling} to reproduce traction forces and shape of keratocytes \\
$k_\textrm{grip}^0$ & Gripping coefficient & 2.5 Pa / (s \um) &Chosen in \cite{shao2012coupling} to reproduce traction forces and shape of keratocytes \\
$k_\textrm{slip}^0$ & Slipping coefficient & 0.25 Pa / \um &Chosen in \cite{shao2012coupling} to reproduce traction forces and shape of keratocytes \\
$r_\textrm{on}$     & Rate of transition from slipping to gripping state & 0.005 s$^{-1}$ &Chosen in \cite{shao2012coupling} to reproduce traction forces and shape of keratocytes \\
$r_\textrm{off}^0$     & Rate of transition from gripping to slipping state (at zero force) & 0.002 s$^{-1}$ &Chosen in \cite{shao2012coupling} to reproduce traction forces and shape of keratocytes \\
$r_\textrm{die}$     & Rate of slipping site death & 0.2 s$^{-1}$ &Chosen in \cite{shao2012coupling} to reproduce traction forces and shape of keratocytes \\
\end{tabular}
\footnotetext[4]{This describes the total number of adhesions over the entire cell.  The value of $N_\textrm{adh}$ in Ref. \cite{shao2012coupling} is listed incorrectly, and should be $N_\textrm{adh}=4000$ over the whole cell.  The change in adhesion number in this paper roughly corresponds to the change in cell area.}

\subsubsection{Numerical evaluation parameters}
\begin{tabular}{lcr}
Parameter & Description & Value \\
\hline
$n \times m$    & Number of (horizontal, vertical) grid points & $256 \times 256$ \\
$L_x \times L_y$ & Box size & $50 \um \times 50 \um$\\
$\Delta t$ & Time step & $2\times 10^{-3} $ s$^*$ \\
$\lambda$ & Cutoff for evaluating phase field equations & $10^{-4}$ \\
\end{tabular}

\subsection{Parameters used for each figure}

\subsubsection{Figure 1}

For the oscillation in Fig. 1, we start with an initial state of a circular cell with radius $6$ \um.  We choose $\eta_m^0 = 61$ pN \um, and have an adhesive stripe of total width $w = 6 \um$, i.e. $\chi(\vb{r}) = \frac{1}{2}\left[1 + \tanh (3 \{\frac{w}{2}-|x|\}/\epsilon)\right]$.  All other parameters are as written in the tables above.  

\subsubsection{Figure 2}

We think of our one-dimensional model as describing a slice down the center of a two-dimensional cell with width $w$, but with $\rho_a$ uniform across the $x$ direction.  Parameters for the actin promoter reaction-diffusion part of the one-dimensional model are exactly the same as for the two-dimensional model of Fig. 1; however, the conservation law follows a slightly different form:
\begin{align}
\int_{-w/2}^{w/2} dx \int_{-L_y/2}^{L_y/2} \left[ \rho_a(y) + \rho_a^\textrm{cyt} \right] \phi(y) = N_a^\textrm{tot}
\end{align}
or, equivalently,
\begin{equation}
\rho_a^\textrm{cyt} = \frac{N_a^{\textrm{tot}}/w - \int d y \rho_a(y) \phi (y)}{\int dy \phi(y)}.  
\end{equation}

The parameters unique to the one-dimensional model are $\alpha = 0.14 \um^3$/s and $\beta = 0.068 \um^3$/s, $m_0 = 2.43 \um^{-2}$ and $\tau = 30 $ s.  The value for $\alpha$ is determined by the sharp interface result, $\alpha = \eta_a^0/4\nu_0$, using the two-dimensional simulation parameters.  We have set $\beta$, $m_0$, and $\tau$ so that the cell oscillates similarly to the two-dimensional simulation.  
The one-dimensional model is evaluated on a grid of $512$ points with $L_y = 100$ \um , with $\Delta t = 0.01$ s.

\subsubsection{Figure 3}

For the turning motion (Fig. 3 top), we start with an initial state of a circular cell with radius $8$ \um.  We choose $\eta_m^0 = 60$ pN \um, and have an adhesive stripe of total width $w = 40$ \um, i.e. $\chi(\vb{r}) = \frac{1}{2}\left[1 + \tanh (3 \{\frac{w}{2}-|x|\}/\epsilon)\right]$.  For the bipedal motion (Fig. 3 bottom), we choose exactly the same parameters, except that we take $w = 10$ \um.  All other parameters are as written in the tables above.

\subsection{Robustness of periodic migration to variation in parameters}

The bulk of our parameters have been set by comparison with experiments on keratocytes, and are identical to those used in \cite{shao2012coupling}; they were not selected to observe periodic migration.  However, some parameters have been changed in order to ensure that the cells polarize and migrate on stripes.  In particular, we changed $N_{a}^{\textrm{tot}}$ and $N_{\textrm{adh}}$ because the cells we study are significantly smaller in area than those in \cite{shao2012coupling}.  We also changed $\eta_m^0$ to change the contraction speed and vary the amplitude of periodic migration.  Initial simulations have shown that periodic migration can be observed over wider ranges of parameters as well; varying one parameter at a time, we see periodic migration at $N_{\textrm{adh}} = 700$, or $\eta_m^0 = 80 $  pN \um, or $N_{a}^{\textrm{tot}} = 400$.  These parameters can be changed more if we change multiple parameters at once.  Our experience with altering the model suggests that periodic migration can be re-created as long as the central polarization mechanism is in place, the contraction and protrusion are closely balanced, and the myosin effectively keeps the memory.  

\section{Details of numerical algorithm}

\subsection{Time-stepping and discretization}

Our goal is to numerically solve the system of equations
\begin{align}
&\partial_t\phi + \textbf{u}\cdot\nabla\phi = \Gamma(\epsilon\nabla^2\phi - G'(\phi)/\epsilon + \epsilon c|\nabla\phi|)\label{eqn_phi}\\
&\partial_t(\phi\rho_{a}) + \nabla\cdot(\phi\rho_{a}\textbf{u}) = \nabla\cdot(\phi D_a \nabla\rho_{a}) + \phi f(\rho_{a},\rho_{a}^{\textrm{cyt}})
 \label{eqn_rhoa}\\
&\partial_t(\phi\rho_{m}) + \nabla\cdot(\phi\rho_{m} \textbf{u}) = \nabla\cdot(\phi D_m(\rho_{a})\nabla\rho_{a}) 
\label{eqn_rhom}
\\
&\nabla\cdot\left[ \nu_0 \phi (\nabla\textbf{u} + \nabla \textbf{u}^T) \right] + \nabla\cdot(\sigma_{\text{poly}}+\sigma_{\text{myo}}) + \textbf{F}_{\text{mem}} + \textbf{F}_{\text{adh}} - \xi\textbf{u} = 0\label{eqn_u}
\end{align}
We fix a uniform spatial grid with grid sizes $\Delta x$, $\Delta y$.  We also use a fixed time step $\Delta t$ to march these equations forward from initial conditions $\phi^{(0)}, \textbf{u}^{(0)},  \rho_{a}^{(0)}, \rho_{m} ^{(0)} $.  We denote the state of the system at time $t = n\Delta t$ by $\phi^{(n)}, \textbf{u}^{(n)},  \rho_{a}^{(n)}, \rho_{m}^{(n)} $.  Suppose we have obtained all these quantities at the time $n \Delta t.$ We then solve all the equations (\ref{eqn_phi})--(\ref{eqn_rhom}) to obtain these quantities at the time $(n+1) \Delta t.$ 

We first obtain $\phi^{(n+1)}$ from the $\phi$-equation (\ref{eqn_phi}) with the forward Euler scheme: 
\begin{align*}
\phi^{(n+1)} = \phi^{(n)} - \Delta t \, \textbf{u}^{(n)}\cdot\nabla\phi^{(n)} + \Delta t \, \Gamma \, \left[\epsilon\nabla^2\phi^{(n)} - G'(\phi^{(n)})/\epsilon + \epsilon c^{(n)}|\nabla\phi^{(n)}| \right] .
\end{align*}
On the right-hand side of this equation, $\nabla\phi^{(n)}$ is calculated with a central difference scheme, $\nabla^2\phi^{(n)}$ is calculated by five-point finite difference scheme, and the curvature term $c^{(n)}$ is calculated by
\[
c^{(n)} = \nabla\cdot\dfrac{\nabla\phi^{(n)}}{|\nabla\phi^{(n)}|}
\]
when $|\nabla\phi^{(n)}|>0.01$, and set to be zero otherwise.

We next solve Eq.\ (\ref{eqn_rhoa}) and Eq.\ (\ref{eqn_rhom}) to obtain $\rho_{a}^{(n+1)}$ and $\rho_{m}^{(n+1)},$ respectively. We apply the forward Euler scheme to the reaction-diffusion-advection equation (\ref{eqn_rhoa}): 
\begin{align*}
\phi^{(n)}\dfrac{\rho_{a}^{(n+1)}-\rho_{a}^{(n)}}{\Delta t} + \dfrac{\phi^{(n+1)}-\phi^{(n)}}{\Delta t} \rho_{a}^{(n)} = -\nabla\cdot(\phi^{(n)}\rho_{a}^{(n)}\textbf{u}^{(n)}) + \nabla\cdot(\phi^{(n)} D_a \nabla \rho_{a}^{(n)}) + \phi^{(n)}f^{(n)}
\end{align*}
Equivalently,
\begin{align}\label{numerical_rhoa}
\rho_{a}^{(n+1)} = \dfrac{(2\phi^{(n)}-\phi^{(n+1)})}{\phi^{(n)}}\rho_{a}^{(n)} - \Delta t \dfrac{\nabla\cdot(\phi^{(n)}\rho_{a}^{(n)}\textbf{u}^{(n)})}{\phi^{(n)}} + \Delta t \dfrac{\nabla\cdot(\phi^{(n)} D_a \nabla \rho_{a}^{(n)})}{\phi^{(n)}} + \Delta t  f^{(n)}
\end{align}
We only divide by $\phi^{(n)}$ in the region where $\phi^{(n)}\ge\lambda$, where $\lambda = 10^{-4}$.  Outside of this region, we keep $\rho_{a}^{(n+1)} = \rho_{a}^{(n)}$. 
More specifically, we have use the following discretization: 
\begin{align*}
\Big[\nabla\cdot(\phi^{(n)}\rho_a^{(n)}\textbf{u}^{(n)}) \Big]_{ij}
&= \left[ \phi^{(n)}_{i+1/2,j} \rho^{(n)}_{{a},i+1/2,j} u^{(n)}_{i+1/2,j} 
- \phi^{(n)}_{i-1/2,j} \rho^{(n)}_{{a},i-1/2,j} u^{(n)}_{i-1/2,j} \right]\Big/\Delta x \\
&+\left[ \phi^{(n)}_{i,j+1/2} \rho^{(n)}_{{a},i,j+1/2} v^{(n)}_{i,j+1/2} - 
\phi^{(n)}_{i,j-1/2} + \rho^{(n)}_{{a},i,j-1/2} v^{(n)}_{i,j-1/2} \right]\Big/\Delta y\\
\Big[\nabla\cdot(\phi^{(n)}D_a\nabla\rho_a^{(n)}) \Big]_{ij}&=D_a\left[ \phi^{(n)}_{i+1/2,j}
\dfrac{\rho_{{a},i+1,j}^{(n)} - \rho_{{a},ij}^{(n)} }{\Delta x}  
-  \phi^{(n)}_{i-1/2,j}  \dfrac{ \rho_{{a},ij}^{(n)} - \rho_{{a},i-1,j}^{(n)} }{\Delta x}  \right]\Big/\Delta x\\
&+    D_a \left[     \phi^{(n)}_{i,j+1/2}  \dfrac{ \rho_{a,i,j+1}^{(n)} - \rho_{a,ij}^{(n)} }{\Delta y}  -  \phi^{(n)}_{i,j-1/2}  \dfrac{ \rho_{a,ij}^{(n)} - \rho_{a,i,j-1}^{(n)} }{\Delta y}  \right]\Big/\Delta y 
\end{align*}
where $\textbf{u}^{(n)}_{ij} = \left(u^{(n)}_{ij}, v^{(n)}_{ij}\right)$. We apply the analogous forward Euler scheme to the $\rho_{m}$-equation (\ref{eqn_rhom}). Since the diffusion coefficient $D_m = D_m(\rho_{a})$ depends on $\rho_{a}$, we discretize the diffusion term at a grid point labeled by $(i,j)$  as follows
\begin{align*}
&\Big[\nabla\cdot(\phi^{(n)}D_m^{(n)}\nabla\rho_{a}^{(n)}) \Big]_{ij} \\
&\quad = \left[    \dfrac{ \phi^{(n)}_{ij}D^{(n)}_{{{m}},ij} + \phi^{(n)}_{i+1,j}D^{(n)}_{{{m}},i+1,j} }{2}   \cdot \dfrac{ \rho_{{{a}},i+1,j}^{(n)} - \rho_{{{a}},ij}^{(n)} }{\Delta x}  -  \dfrac{ \phi^{(n)}_{ij}D^{(n)}_{{{m}},ij} + \phi^{(n)}_{i-1,j}D^{(n)}_{{{m}},i-1,j} }{2}   \cdot \dfrac{ \rho_{{{a}},ij}^{(n)} - \rho_{{{a}},i-1,j}^{(n)} }{\Delta x}  \right]\Big/\Delta x\\
&\quad +  \left[    \dfrac{ \phi^{(n)}_{ij}D^{(n)}_{{{m}},ij} + \phi^{(n)}_{i,j+1}D^{(n)}_{{{m}},i,j+1} }{2}   \cdot \dfrac{ \rho_{{{a}},i,j+1}^{(n)} - \rho_{{{a}},ij}^{(n)} }{\Delta y}  -  \dfrac{ \phi^{(n)}_{ij}D^{(n)}_{{{m}},ij} + \phi^{(n)}_{i,j-1}D^{(n)}_{{{m}},i,j-1} }{2}   \cdot \dfrac{ \rho_{{{a}},ij}^{(n)} - \rho_{{{a}},i,j-1}^{(n)} }{\Delta y}  \right]\Big/\Delta y 
\end{align*}
where $D_{m, ij}^{(n)} = D_m (\rho^{(n)}_{{a},ij}).$
To keep $\rho_{m}$ conserved and reduce its drift, we rescale $\rho_m$ at each time step so that the total integral of $\rho_{m}$ is kept a constant. We note that we have corrected the position of the non-constant diffusion coefficient in the $\rho_{m}$ equation in \cite{shao2012coupling}.

Finally, we solve the Stokes equation (\ref{eqn_u}) with a semi-implicit Fourier spectral scheme to obtain $\textbf{u}^{(n+1)}.$  
To do so, we first subtract the term $\nu_0 \tilde{\phi} \nabla^2 \textbf{u}$ from both sides of the Stokes equation (\ref{eqn_u}) with $\tilde{\phi}$ a constant (e.g., $\tilde{\phi} = 2$)
to yield
\begin{align*}
\xi\textbf{u} - \nu_0 \tilde{\phi} \nabla^2 \textbf{u}  &= \nabla\cdot\left[ \nu_0 (\phi-\tilde{\phi}) \nabla\textbf{u} + \nu_0 \phi \nabla \textbf{u}^T) \right] + \nabla\cdot(\sigma_{\text{poly}}+\sigma_{\text{myo}}) + \textbf{F}_{\text{mem}} + \textbf{F}_{\text{adh}} 
\equiv \textrm{RHS}(\textbf{u},\phi,\rho_{a},\rho_{m})
\end{align*}
To obtain $\textbf{u}^{(n+1)},$ we set $\textbf{u}^{(n+1)}_0 = \textbf{u}^{(n)}$ and solve the following equation iteratively using the spectral Fourier method: 
\begin{align*}
\xi\textbf{u}^{(n+1)}_{k+1} - \nu_0 \tilde{\phi} \nabla^2 \textbf{u}^{(n+1)}_{k+1} 
= \textrm{RHS}(\textbf{u}^{(n+1)}_k,\phi^{(n+1)},\rho_{a}^{(n+1)},\rho_{m}^{(n+1)})
\qquad k = 0, 1,  \dots, m  
\end{align*}
and set $\textbf{u}^{(n+1)}  = \textbf{u}^{(n+1)}_m.$ 
The calculations of $\nabla\cdot(\sigma_{\text{poly}}+\sigma_{\text{myo}}),\textbf{F}_{\text{mem}}$ and $\textbf{F}_{\text{adh}}$ are performed as in \cite{shao2012coupling}.  
The number of steps $m$ in this iteration is set to be $m = 10$ or set by 
\begin{align*}
\max | \textbf{u}^{(n+1)}_{m} - \textbf{u}^{(n+1)}_{m-1} | < 0.01 \; \textrm{max} |   
 \textbf{u}^{(n+1)}_{m} |.
\end{align*}

Shifting of the simulation box when the cell approaches the box edges is performed as in \cite{shao2012coupling,shao2010computational}.  

\subsection{Adhesion dynamics and calculation of adhesion force}

 The adhesion dynamics are precisely as given in \cite{shao2012coupling}, except that adhesions do not form off of the adhesive stripe, and are destroyed if they leave the stripe.  For completeness, we summarize these dynamics here.

Adhesions between the cell and substrate are tracked individually; there are a fixed number $N_{\textrm{adh}}$ of adhesions, and if one is destroyed, another one is created.  The probability of adhesion formation is proportional to $\rho_a$ and to $\phi$, resulting in nascent adhesions being more likely to form at the front of the cell.  We compute the initial adhesion location by a rejection method: we propose an adhesion location $\rb_0$ distributed uniformly in the region $|x|\le\frac{w}{2}$, where $w$ is the total width of the adhesive stripe, and accept that adhesion location with probability $p = \rho_a(\rb_0) \phi / \textrm{max}(\rho_a)$.  Adhesions are destroyed if they leave the stripe (i.e. have $|x|>\frac{w}{2}$) or if $\phi < 1/2$ at the adhesion location.  

Adhesions are advected by the cytoskeletal flow, $\uv$; in practice, we choose the adhesion velocity to be the velocity $\uv$ at the nearest grid point to the adhesion location.

Adhesions have two modes: ``slipping'' and ``gripping.''  Adhesions are formed in gripping mode.  When an adhesion is formed or transitions into gripping mode, its initial position $\rb_0$ is noted.  The gripping adhesion acts as a spring stretched from its initial location (where the adhesion attaches to the substrate) to its current location.  It thus exerts a force on the cell of
\begin{equation}
\vb{F}_{\textrm{grip}} = -k_{\textrm{grip}} (\rb - \rb_0)
\end{equation}
By contrast, a slipping adhesion exerts a force 
\begin{equation}
\vb{F}_{\textrm{slip}} = -k_{\textrm{slip}} \uv(\rb)
\end{equation}
where $\rb$ is the adhesion position.  We assume that the adhesions mature over time: $k_\textrm{grip} = k_\textrm{grip}^0 t_{\textrm{adh}}$ and $k_\textrm{slip} = k_\textrm{slip}^0 t_{\textrm{adh}}$ where $t_{\textrm{adh}}$ is the age of the adhesion site.  
Adhesions may transition between slipping and gripping, and slipping adhesions may disappear.  Gripping adhesions rupture and become slipping adhesions with a force-dependent rate $r_{\textrm{off}} = r_{\textrm{off}}^0 \exp(|\vb{F}_\textrm{grip}|/F_0)$, with $F_0$ the gripping strength scale.  Slipping adhesions may return to gripping mode with a rate $r_{\textrm{on}}$, and disappear with a rate $r_{\textrm{die}}$.  To calculate the force density $\vb{F}_{\textrm{adh}}$ that enters into the Stokes equation, forces on adhesions are distributed to the nearest grid point; we therefore list the appropriate units in terms of forces per unit area.  (We note that \cite{shao2012coupling} incorrectly describes the force as being spread over the closest four grid points.)  



\begin{thebibliography}{55}
\bibitem [{myo()}]{myosin_note}%
  \BibitemOpen
  \href@noop {} {}\bibinfo {note} {Myosin is swept to the cell rear by the
  retrograde cytoskeletal flow, as $\uv$ is less than the cell velocity through
  most of the cell; see \cite{shao2012coupling}.}\BibitemShut {Stop}%
\end{thebibliography}

\end{document}